\documentclass[doublecol,figures]{epl2} 
\def\LiV{LiVCuO$_4$}
\def\beq{\begin{equation}}
\def\eeq{\end{equation}}

\title{Multiferroicity due to nonstoichiometry  in  the chain cuprate LiVCuO$_4$}
\shorttitle{Multiferroicity due to nonstoichiometry} 

\author{A.S.\ Moskvin\inst{1} \and S.-L.\ Drechsler\inst{2} }
\shortauthor{A.S.\ Moskvin and S.-L.\ Drechsler}

\institute{                    
  \inst{1} Ural State University, 620083 Ekaterinburg,  Russia\\
  \inst{2} Leibniz Institut f\"ur Festk\"orper- und Werkstoffforschung
Dresden, D-01171, Dresden, Germany
}

\abstract{
The recently  observed multiferroic behavior in the 
$s$=1/2 chain cuprate LiVCuO$_4$ ($\equiv$ LiCuVO$_4$)  with edge-shared CuO$_4$ plaquettes and 
helical spin ordering does not agree with the existing theories such as a spin-current scenario.  We argue that the effect can be consistently 
explained, if  the exchange-induced electric polarization on the
out-of-chain Cu$^{2+}$ centers substituting for Li-ions in LiVCuO$_4$
is taken into account. These substituent centers are proved to be an 
effective probe of the spin incommensurability and the magnetic field effects.}

\begin{document}

\maketitle

 
The recent observations of multiferroic behaviour in edge-shared chain cuprates LiVCuO$_4$ 
\cite{Naito,Naito1} and LiCu$_2$O$_2$   
\cite{Cheong} 
challenge the multiferroic community. At 
first sight, these quantum $s$=1/2 1D helicoidal magnets seem to  be  
prototypical examples of 1D spiral-magnetic ferroelectrics revealing the  
"ferroelectricity caused by spin-currents"\cite{Katsura1} 
with the textbook expression for the $uniform$ polarization induced by a 
spin spiral    with the wave vector $\bf Q$: 
\beq
{\bf P}\propto\left[{\bf e}_3\times {\bf Q}\right],
\label{PM}
\eeq
 where ${\bf e}_3$ is a vector orthogonal to the spin spiral plane\cite{Mostovoy}. 
However, both \LiV \, and LiCu$_2$O$_2$ show up a behavior which is obviously counterintuitive within the framework of spiral-magnetic ferroelectricity\cite{Naito1,Cheong}. Indeed, the spontaneous ferroelectric polarisation in LiVCuO$_4$ is observed along the $a$-axis in agreement with the spin-current scenario.  $P_a$ decreases with increasing external field along $a$-axis and vanishes for $h>2$ T, where the $bc$-plane helical structure is realized, however, without any signatures of the ferroelectricity along $c$-axis predicted by the existing theories.
 The saturated value of $P_a$ in LiVCuO$_4$ in $h=0$ reveals a magnitude ($P_a\approx 43\mu C/m^2$) comparable with that of the 
multiferroic Ni$_3$V$_2$O$_8$ where the 
magnetic ordering drives the electric polarization $\approx 10^{2}\mu C/m^2$ (ref.\cite{Lawes}) that represents a typical magnitude of polarization induced by magnetic reordering in multiferroics. However, such an anomalously strong magnetoelectric effect seems to be an unexpexted feature for a
system with $e_g$-holes and a nearly
perfect  highly symmetric chain structure (see Fig. 1)  
\begin{figure}[t]
\includegraphics[width=7.5cm,angle=0]{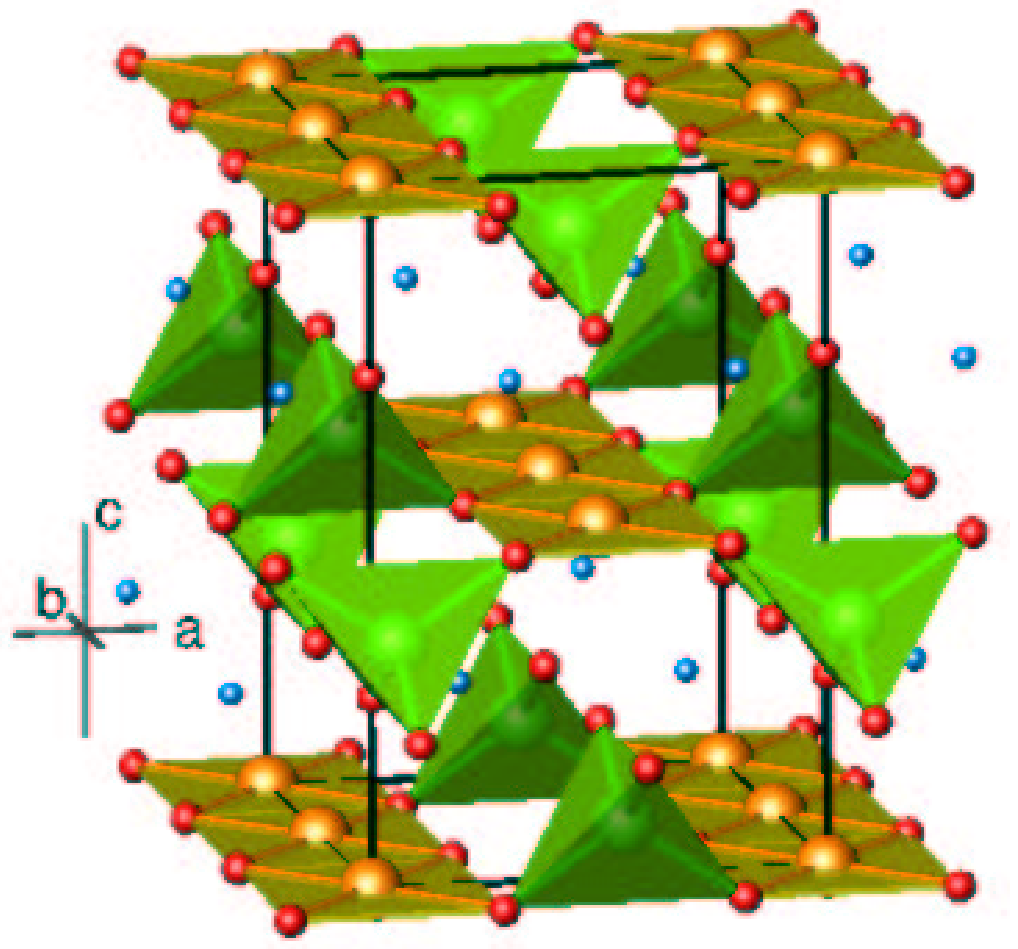}
\caption{(Color online). Crystal structure of LiVCuO$_4$. (In the on-line colour version, the O$^{2-}$ ions are highlighted in red, the Li$^{+}$ ions  in blue, the VO$_4$ tetrahedra are highlighted in green, the CuO$_2$ chains in light olive).} \label{fig1}
\end{figure}  
which both are  unfavourable  for a strong spin-electric coupling. 
Thus the  magnetoelectric effect  in the title cuprate \LiV \, raises 
fundamental questions about its microscopic origin.
 
Currently  two essentially different spin structures of net electric polarization in crystals are considered: 1) with a bilinear {\it nonrelativistic symmetric} spin coupling\cite{Chapon,Betouras,Sergienko1}  
\begin{equation}
	{\bf P}_s=\sum_{mn}{\bf \Pi}_{mn}^s({\bf S}_m\cdot {\bf S}_n)\,
	\label{TMS}
\end{equation}
or 2) a bilinear {\it relativistic antisymmetric} spin coupling\cite{Sergienko,Katsura1,Mostovoy}  
\begin{equation}
	{\bf P}_a=\sum_{mn}\left[{\bf \Pi}_{mn}^a\times \left[{\bf S}_m\times {\bf S}_n\right]\right]\, ,
\end{equation}
respectively. The effective dipole moments ${\bf \Pi}_{mn}^{s,a}$  depend on the $m,n$ orbital states and the $mn$ bonding geometry.  First term  stems somehow or other from a spin isotropic Heisenberg exchange interaction, see, e.g. ref.\cite{TMS}, and is not at work for CuO$_2$ chains in \LiV \, because of inversion symmetry consideration. 
 The second, or "spin-current" term is at present frequently considered to be 
one of the 
leading mechanisms of multiferroicity\cite{Sergienko,Katsura1,Mostovoy,Katsura2,Jia,Hu}. However, there 
are notable exceptions, in particular the manganites RMn$_2$O$_5$, HoMnO$_3$, where  a ferroelectric polarization can appear without any indication of magnetic chiral symmetry breaking\cite{Chapon,Sergienko1}, and delafossite CuFe$_{1-x}$Al$_x$O$_2$, where the helimagnetic ordering generates a spontaneous electric polarization $\parallel$ to the helical axis\cite{CuFeAlO}, in sharp contrast with the prediction of the spin current model. 
  At closer examination the spin current mechanism comes to a 
dipole polarization of the three-center M$_1$-O-M$_2$ system due to an exchange-relativistic antisymmetric Dzyaloshinsky-Moriya (DM) type coupling with or 
without lattice degrees of freedom involved \cite{Sergienko,Katsura1}. 
However, the original "spin-current" model by Katsura {\it et al.}\cite{Katsura1} and its later versions \cite{Jia,Hu} seem to be  questionable  as the authors proceed with an unrealistic scenario  \cite{elsewhere}.
 At variance with such a scenario one can apply a conventional procedure to derive an effective {\it spin-operator} ${\bf \hat P}_a=\left[{\bf \Pi}_{12}^a\times \left[{\bf \hat S}_1\times {\bf \hat S}_2\right]\right]$  for the electric dipole moment in the three-center M$_1$-O-M$_2$ system like  a technique suggested in refs.\cite{DM-JETP,DM-PRB} to derive 
expressions for the Cu and O spin-orbital contributions to 
the Dzyaloshinsky-Moriya coupling in cuprates. However, a straightforward application of this technique to the system of the $e_g$-holes in a  perfect chain structure 
of edge-shared CuO$_4$ plaquettes as in \LiV \, shows that the 
spin current does not produce an electric polarization\cite{elsewhere}. 
An alternative mechanism of giant magnetoelectricity based on the 
antisymmetric DM type magnetoelastic coupling was proposed recently
\cite{Sergienko}. However, here we meet with a "weak" contributor. Indeed, the 
minimal value of $\gamma$ parameter ($\gamma = d{\bf D}/d{\bf R}$) needed to 
explain the experimental  phase transition in multiferroic manganites 
is two orders of magnitude larger than 
reasonable microscopic estimates\cite{Sergienko}.
 Thus we may state that the ideal edge-shared CuO$_4$ plaquettes  chain 
structure
appears to be robust regarding the proper spin-induced electric polarization. It means that we should look for the origin of the puzzling multiferroicity in \LiV \, somewhere within the out-of-chain stuff.

In the Letter we argue that 
 the ferroelectricity in  this compound is actually {\it unrelated}  to spin currents. 
It can be consistently explained if one takes into account the $nonrelativistic$ effects of the spin-dependent electric polarisation \cite{elsewhere} due to the out-of-chain Cu$^{2+}$ centers substituting for Li-ions in LiVCuO$_4$ present in real samples. For an illustration we start with  a short description of a simple   theory 
of the exchange-induced electric polarisation (\ref{TMS}) which generalizes the model approach by Tanabe {\it et al.}\cite{TMS}.  
For simplicity, let us consider an one-particle (electron/hole) center in a crystallographically centrosymmetric position of a magnetic crystal. Then all the particle states exhibit a definite spatial parity, even (g) or odd (u), respectively. Having in mind the 3d centers, we'll assume an
 even-parity ground state $|g\rangle$. For simplicity we restrict ourselves by only 
one excited odd-parity state  $|u\rangle$. An isotropic exchange coupling with the surrounding spins can be written as follows:
\begin{equation}
	{\hat V}_{ex}=\sum_n{\hat I}({\bf R}_n)({\bf s}\cdot {\bf S}_n),
\end{equation}
where ${\hat I}({\bf R}_n)$ is an orbital operator with a matrix
\begin{math}
\left(\begin{array}{clrr}  I_{gg}({\bf R}_n) & I_{gu}({\bf R}_n) \\
I_{ug}({\bf R}_n) & I_{uu}({\bf R}_n)
\end{array}\right)\, .
\end{math}
The parity-breaking off-diagonal part of exchange coupling can lift the center of symmetry and mix $|g\rangle$ and $|u\rangle$ states 
giving rise to a nonzero electric dipole polarization of the ground state
\begin{equation}
{\bf P}=2c_{gu}\langle g|e{\bf r}|u\rangle =\sum_{n}{\bf \Pi}_n({\bf s}\cdot {\bf S}_n)\, ,
\end{equation}
where 
\begin{equation}
{\bf \Pi}_n =	2I_{gu}({\bf R}_n)\frac{\langle g|e{\bf r}|u\rangle}{\Delta_{ug}} \, .
\end{equation}
and $\Delta _{ug}=\epsilon_u-\epsilon_g$. 

It should be noted that at variance with the DM type, or the spin current  
scenario the direction of the exchange-induced dipole moment for $i,j$ pair does not depend on the direction of the spins ${\bf S}_i$ and ${\bf S}_j$. In other words, the spin-correlation factor $\langle{\bf S}_i\cdot {\bf S}_j\rangle$ modulates a pre-existing dipole moment ${\bf \Pi}_{ij}$ which direction and value depend on the Me$_i$-O-Me$_j$ bond geometry and the orbitals involved in the exchange coupling. 
The net exchange induced polarization of the magnetic crystal depends both on the crystal symmetry and the spin structure. 

The orthorhombicaly distorted inverse spinel structure of  
LiVCuO$_4$ \cite{V-Cu} contains chains of edge-shared LiO$_6$ and CuO$_6$ octahedra running along ${\bf a}$ and ${\bf b}$ crystal axes, respectively \cite{Prokofiev}. The two neighboring CuO$_2$ chains within the same $ab$ plane are connected by VO$_4$ tetrahedra that alternate up and down along the chain direction, while two CuO$_2$ chains within the same unit cell are connected by LiO$_6$ octahedra (see Fig.\,\ref{fig1})
First neutron-diffraction measurements 
\cite{Gibson} 
and a more detailed INS study 
\cite{Enderle} have revealed 
a long-range $incommensurate$ helical magnetic ordering with a 
propagation vector ${\bf q}=(0, 0.532, 0)$ and moments confined to 
the $ab$-plane below $T_N\approx 2.1$ K., i.e.\
the spin spiral runs along the $b$-axis: 
${\bf S}(y)= S(\cos(\theta ),\sin(\theta ),0)$ 
is
the classical ground state of an isolated Heisenberg chain 
in \LiV. Here  $\theta=qy+\alpha$,   $\alpha$ is a phase shift. 
 The two CuO$_2$ chains within the same unit cell are 
ordered antiferromagnetically: i.e.\ the phase shift $\Delta \alpha = \pi$. 
A combined NMR and AFMR study\cite{Butgen} also shows that in the 
low-field range 
H $<$ H$_{c1}\approx$ 2.5 T, an $ab$-oriented planar spiral spin structure is 
realized 
in agreement with 
refs.\ 23,24. Based on NMR spectra simulations\cite{Butgen} and neutron scattering studies\cite{Naito1}, the 
transition at H$_{c1}$ can be 
attributed to a spin-flop transition, where the spin plane of 
the magnetically ordered structure rotates to be 
perpendicular to the direction of the magnetic field  applied in the easy plane.

Noteworthy,  LiVCuO$_4$ single crystals can reveal 
a sizeable non-stoichiometry with Li deficiency and appearance 
of Cu$^{2+}$ or Cu$^{3+}$ ions on 
native Li ($4d$) positions\cite{Prokofiev} due to similar
ionic radii of Cu and Li. Indeed, a valence bond analysis 
\cite{Brown} for such a Cu center
with unrelaxed nearest neighbor O distances of $r_1$=2.1 \AA \ 
and $r_2$=2.18 \AA \
yields a valency $\nu_{Cu}=\sum s_{Cu-O} \approx$ 2:
$$\nu_{Cu}=4\exp\left(\frac{r_0-r_1}{B}\right)+2\exp
\left(\frac{r_0-r_2}{B}\right)=1.798,$$
where $r_0(Cu^{2+})$=1.679 \AA \ and $B=0.37$\AA .
  Adopting $r_1$=2.05 \AA \ for the Cu-O distances near the basal plane 
after a small relaxation of the adjacent unshared O ions
above/below the V ions towards the Cu center
one obtains $\nu_{Cu}$=1.98. These numbers should be compared 
with $\nu_{Cu} = 2.16$ obtained for a regular Cu chain site. 

Significant, up to 5\% concentration of Cu ions on Li-sites was assumed by Kegler  {\it et al.}\cite{Kegler} to explain the appearance of an additional weak $^7$Li NMR signal. The Li ions reside in two types of the centrosymmetrical positions A,B that differ by a bc-plane mirroring (see Fig.\,1). Each Li ion has four nearest Cu$^{2+}$ ions at the same distance. 
Below we argue that  the Cu ions substituting for Li positions form  strongly polarizable 
entities whose electric polarization due to a parity-breaking exchange interaction 
with the spin spirals in the chains is responsible for the multiferroicity in \LiV.

\begin{figure}[t]
\includegraphics[width=8.0cm,angle=0]{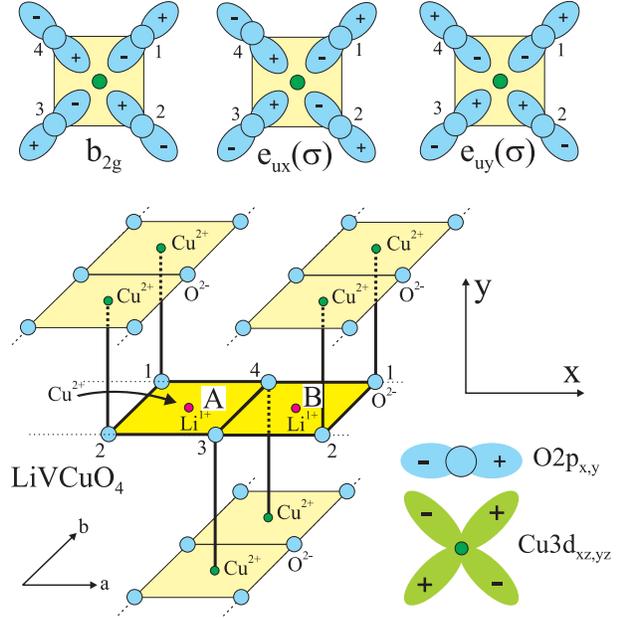}
\caption{(Color online). A slightly idealized view of the Li-O-Cu bonds in the unit cell of LiVCuO$_4$. An impurity center with a 
Cu-ion substituted for Li is inbetween upper and lower CuO$_2$ chains from the same unit cell.  Top picture illustrates the hole density distributions for active even- and odd-parity O 2$p$ orbitals. Main hole p-d transfer channel is shown.
} \label{fig2}
\end{figure}
\begin{figure}[t]
\includegraphics[width=8.0cm,angle=0]{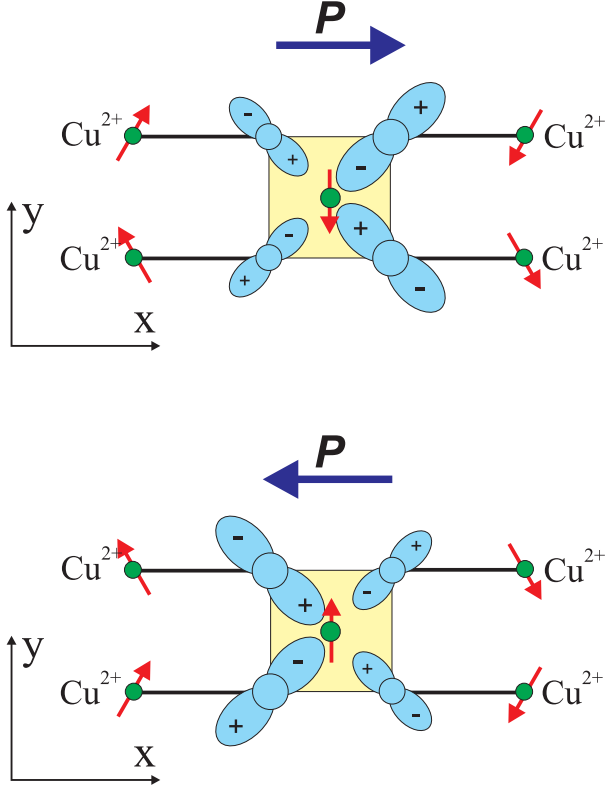}
\caption{(Color online). Illustration of the exchange-induced polarisation effect for 
a Cu 3$d$ O2$p$ $b_{2g}$ hole at a Li-position of LiVCuO$_4$. 
The chain spiral Cu$^{2+}$ spins are arranged to give nonzero G$_x$ and A$_y$ modes.} \label{fig3}
\end{figure}

 
  Fig.\,\ref{fig2} shows both crystalline surrounding of the Cu$^{2+}$ substituent center and oxygen-centered hole density distribution for active even-parity $b_{2g}(\propto d_{xy})$ and odd-parity $e_u(\sigma)$ orbitals within the D$_{4h}$ point group.  
  Let us consider the exchange-induced electric polarisation effects for these centers starting from a somewhat idealized "purely oxygen" model  with active O2p $b_{2g}$ and $e_{u}(\sigma)$ orbitals exchange-coupled with two CuO$_2$ chains (see Fig.\,\ref{fig2}).
 These orbitals are coupled by a large electric dipole matrix element:
 \begin{equation}
\langle b_{2g}|e{\bf x}|e_{uy}(\sigma)\rangle	=\langle b_{2g}|e{\bf y}|e_{ux}(\sigma)\rangle	=\frac{1}{\sqrt{2}}eR_{CuO} \, ,
\end{equation}
where $R_{CuO}\approx$ 2 \AA \ is the Cu-O distance.  
   Main exchange hole-hole coupling is determined by a "vertical" pd$\pi$ transfer between O2px,y orbitals of central plaquette and Cu 3d$_{xz,yz}$ ($e_g$-) orbitals of Cu ions in upper and lower chains. Diagonal $I_{gg}=I_{b_{2g}b_{2g}}$ and off-diagonal $I_{gu}=I_{b_{2g}e_{u}}$ exchange integrals for e.g. Cu$_1^{2+}$-A bond can be written as follows:
$$
	I_{gg}(Cu_n)= t_{pd\pi}^2\left(\frac{1}{\Delta _{e_{g}}^{s}} -\frac{1}{\Delta _{e_{g}}^{t}}\right) \,  ,
$$   
\begin{equation}
I_{gux}(1)=I_{guy}(1)=
\end{equation}
$$
\frac{1}{2}t_{pd\pi}^2\left(\frac{1}{\Delta _{e_{g}}^{s}} -\frac{1}{\Delta _{e_{g}}^{t}}+\frac{1}{\Delta _{e_{g}}^{s}-\Delta _{ug}} -\frac{1}{\Delta _{e_{g}}^{t}-\Delta _{ug}}\right)\, ,
$$
where
 $\Delta _{e_{g}}^{s}=\epsilon _{e_{g}}+A+B+2C$, $\Delta _{e_{g}}^{t}=\epsilon _{e_{g}}+A-5B$ are the energies of the two-hole copper singlet and triplet terms 
with  $b_{2g}e_{g}$ configurations, respectively, A, B, and C are Racah 
parameters, $\Delta _{ug}=\epsilon _{e_{u}}-\epsilon _{b_{2g}}$. The exchange channel we consider results in a significant ferromagnetic $gg$-coupling between Cu$^{2+}$-substituent and adjacent CuO$_4$-centers from lower and upper chains. However,  
 the magnitude of the off-diagonal $gu$-exchange integrals can sufficiently exceed that of a conventional diagonal exchange integral mainly due to a smaller value of the energy 
separation $(\Delta _{e_{g}}^{s,t}-\Delta _{ug})$ as compared with $\Delta _{e_{g}}^{s,t}$.
Simple symmetry considerations point to the following relations between exchange-dipole parameters ${\bf \Pi}_n$
$$
\Pi_{1x}=\Pi_{2x}=-\Pi_{3x}=-\Pi_{4x}=\pm \Pi\, ;
$$
\begin{equation}
\Pi_{1y}=-\Pi_{2y}=-\Pi_{3y}=\Pi_{4y}=\Pi \, ,
\end{equation}
where upper and lower signs correspond to positions A and B, respectively.
Illustration of the exchange-induced polarization effect for O2p $b_{2g}$ hole in Li-position of LiVCuO$_4$ is presented in Fig.\,\ref{fig3}. It should be emphasized that the induced electric dipole moment is of purely electronic nature and lies in the $ab$ plane.
To describe different configurations of the spin neighbourhood for 
a Cu$^{2+}$ substituent  (see Fig.\,\ref{fig2}),
 we introduce four basic vectors similarly to conventional ferro- and antiferromagnetic vectors as follows:
$$
{\bf F}(y)=[{\bf S}_1	+{\bf S}_2+{\bf S}_3+{\bf S}_4]; \, 
{\bf G}(y)=[{\bf S}_1	-{\bf S}_2+{\bf S}_3-{\bf S}_4];
$$
$$
{\bf A}(y)=[{\bf S}_1	+{\bf S}_2-{\bf S}_3-{\bf S}_4]; \,
{\bf C}(y)=[{\bf S}_1	-{\bf S}_2-{\bf S}_3+{\bf S}_4] \, 
$$
($y$ points to the substituent coordinate along a spin chain) with a kinematic constraint: $({\bf F}\cdot{\bf A})$=$({\bf C}\cdot{\bf G})$=0 valid for two identical spirals in the unit cell irrespective of phase shift. In terms of new vectors we arrive at simple expressions for local electric polarization in A and B positions:
\begin{equation}
P_a(y)=\pm d_a({\bf s}\cdot {\bf A})\, ; \, P_b(y)=d_a({\bf s}\cdot {\bf C})\, .	
\end{equation}
For two ferromagnetically ordered spirals (phase shift $\Delta\alpha =0$) ${\bf A}={\bf G}=0$, while for two antiferromagnetically ordered spirals as in \LiV \, (phase shift $\Delta\alpha =\pi $) ${\bf F}={\bf C}=0$. 
It means that  the  $b$-component of the electric polarisation $P_b(y)$ for a mutual antiferromagnetic (antiphase)   spin ordering of the "upper" and the 
"lower" CuO$_2$ chains (${\bf S}_1	=-{\bf S}_4, {\bf S}_2=-{\bf S}_3$) vanishes: $P_b(y)=0$ irrespective of the spin-spiral plane orientation, and the substituent electric polarisation may be oriented only along crystal $a$-axis. Moreover, the nonzero net crystal polarisation $\left\langle P_a(y)\right\rangle $ one obtains only, if ${\bf s}_A\not={\bf s}_B$, or for magnetically nonequivalent A and B substituent positions. 
Spin polarisation of the Cu$^{2+}$ substituent spin can be easily found within the framework of a weak coupling approximation, if one takes the most general form of the impurity-spiral ground state ($gg$) exchange interaction
\beq
V_{sS}=\sum_{i=1-4}\hat{\bf s}{\bf \stackrel{\leftrightarrow}{I}(i)} \,\hat{\bf S}_i =(\hat{\bf s}\cdot \hat{\bf H}_0)\, ,\label{1}
\eeq
where $\hat{\bf H}_0$ is an effective magnetic field, acting on the Cu$^{2+}$ substituent, $I_{\alpha\alpha}(i)=I_{\alpha\alpha},I_{xz}(i)=\pm I_{xz};
I_{xy}(1)=-I_{xy}(2)=I_{xy}(3)=-I_{xy}(4)=\pm I_{xy};
I_{zy}(1)=-I_{zy}(2)=I_{zy}(3)=-I_{zy}(4)=I_{zy}
$
form a symmetric matrix of the exchange integrals, the upper and lower signs correspond to positions A and B, respectively. Thus for effective field we obtain
\begin{equation}
{\bf H}_0(y)={\bf \stackrel{\leftrightarrow}{I}_F}{\bf F}+{\bf \stackrel{\leftrightarrow}{I}_G}{\bf G}
\end{equation}
with
$$
{\bf \stackrel{\leftrightarrow}{I}_F} =\pmatrix{I_{xx} & 0 & \pm I_{xz} \cr 0 & I_{yy} & 0 \cr \pm I_{xz} & 0 & I_{zz}\cr}; 
{\bf \stackrel{\leftrightarrow}{I}_G}=
\pmatrix{0 & \pm I_{xy} & 0 \cr \pm I_{xy} & 0 & I_{zy} \cr 0 & I_{zy} & 0\cr}.
$$
Effective molecular field on the Cu$^{2+}$ substituent spin in \LiV \, is determined solely by anisotropic terms due to a complete cancellation of isotropic contributions (${\bf F}=0$).  
Thus, the exchange-induced electric polarisation $P_a(y)$ for O2p $b_{2g}$ hole in Li-position can be written as follows:
 $$
P_a(y)=\pm d_a({\bf s}\cdot {\bf A})= \pm \frac{d_aS}{H(y)}({\bf h}+{\bf H}_0)\cdot {\bf A} =\pm \frac{d_aS}{H(y)}({\bf h}\cdot {\bf A}+
$$
\beq
\pm I_{xy}(G_xA_y+G_yA_x)+I_{zy}(G_zA_y+G_yA_z))\, ;
\label{Pa}
\eeq
where $d_a=\sqrt{2}eR_{CuO}\Delta _{ug}^{-1}I_{b_{1g}e_{ux}(\sigma)}(Cu_1)$, ${\bf h}$ is an external magnetic field. It is worth noting that the effective magnetic field in these expressions reflects the symmetry of conventional ($gg$) exchange coupling, while the basic spin vectors ${\bf A}$ and ${\bf C}$ do that of the parity-breaking ($ug$) exchange coupling.

We start with an $ab$-planar spiral 
as the zero-field ground state of the CuO$_2$ chains in \LiV \, and assume T=0. 
Then the  nonzero basic spin vectors and the resultant magnetic field 
acting on the spin of the Cu$^{2+}$ 
substituent at $y$-position can be written as follows:
$$
G_x=-4S\sin(\frac{qb}{2})\sin(qy);G_y=4S\sin(\frac{qb}{2})\cos(qy);
$$
\beq
A_x=4S\cos(\frac{qb}{2})\cos(qy);A_y=4S\cos(\frac{qb}{2})\sin(qy).
\eeq 
$$
H_x=h_x+H_{0x}=h_x\pm 4SI_{xy}\sin(\frac{qb}{2})\cos(qy);
$$
$$
H_y=h_y+H_{0y}=h_y\mp 4SI_{xy}\sin(\frac{qb}{2})\sin(qy);
$$
\beq
H_z=h_z+H_{0z}=h_z+4SI_{zy}\sin(\frac{qb}{2})\cos(qy)\, ,
\eeq
We see that both all the components and absolute value of effective magnetic field  oscillate along the chain direction in a complex way thus providing at variance with the DM type mechanism the $oscillatory$ behavior of electric polarisation accompanying the spin spiral ordering. 
\begin{figure}[t]
\includegraphics[width=7.0cm,angle=0]{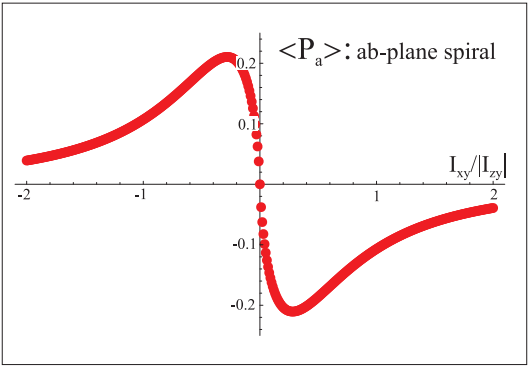}
\caption{(Color online). The dependence of  the mean value of the $a$-axis electric polarisation of the Cu$^{2+}$ substituent $\left\langle P_a(y)\right\rangle$ (in units of $d_a/$ion) on the  ratio of exchange parameters $I_{xy}/I_{zy}$.}  \label{fig4}
\end{figure} 
 \begin{figure}[t]
\includegraphics[width=7.5cm,angle=0]{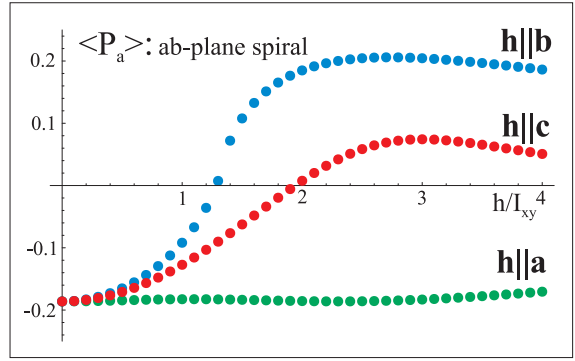}
\caption{(Color online). The  field dependence of the mean value of the $a$-axis electric polarisation of the Cu$^{2+}$ substituent $\left\langle P_a(y)\right\rangle$ (in units of $d_a/$ ion) for $ab$-plane spiral:  $I_{xy}/I_{zy}=0.5$. }  \label{fig5}
\end{figure}
Electric polarisation $P_a(y)$ oscillates along 
the chain direction $b$  as follows:
$$
P_a(y)=\pm\frac{2d_aS \cos(\frac{qb}{2})}{H(y)}[h_x\cos(qy)+h_y\sin(qy)
$$
\beq
\pm 2I_{xy}\sin(\frac{qb}{2})\cos(2qy)]\, .
\label{P_ab}
\eeq
Interestingly, the change of the spiral modulation wave length, 
inversely proportional to
 the pitch angle ($qb$), should affect the magnitude of the electric polarisation.
Formally, the spontaneous ($h=0$) zero-temperature polarisation $P_a(y)$ does  depend on the $I_{xy}/|I_{zy}|$ ratio rather than on the numerical values of exchange anisotropy parameters. It  changes sign with that of $I_{xy}$. At the same time, $\left\langle P_a(y)\right\rangle \rightarrow $0 given $I_{zy}\rightarrow $0. In Fig.\,\ref{fig4} we presented the dependence of the mean value of the $a$-axis electric polarisation of the Cu$^{2+}$ substituent $\left\langle P_a(y)\right\rangle$ (in units of $d_a/$ion) given $qb/2=0.233\pi$ that corresponds to a pitch angle $\approx 84^{\circ}$\cite{Gibson,Enderle}) on the $I_{xy}/I_{zy}$ ratio that points to a significance of both integrals in the magnetoelectric effect.
 
We see that for a reasonable 
ratio between the exchange integrals $I_{xy}$ and $I_{zy}$ 
the averaging procedure can significantly reduce the local polarisation 
from several times to more than one order of magnitude. Maximal magnitudes of polarisation one obtains for $|I_{xy}/I_{zy}|\approx 0.4$. 
Fig.\,\ref{fig5} shows the field dependence of the mean value of the $a$-axis electric polarisation of the Cu$^{2+}$ substituent $\left\langle P_a(y)\right\rangle$ in frames of a rigid $ab$-plane spiral approximation given $I_{xy}/I_{zy}=0.5$. In other words, we assume the external magnetic field does not distort the spin spiral, however, polarises the substituent spin. It is worth noting that the 
averaging procedure should be performed carefully, as we deal with a very slow convergence of the chain summing. 

For the $bc$-plane spin spiral ordering the electric polarisation $P_a(y)$ oscillates both for A and B substituent positions with a magnitude which can be easily obtained from (\ref{P_ab}), if one 
makes the  interchange:$h_x\rightarrow h_z$, $\pm I_{xy}\leftrightarrow I_{zy}$. In other words, it means that in a zero magnetic field $P_a(I_{xy}/I_{zy})|_{ab}=P_a(I_{zy}/I_{xy})|_{bc}$.  However, at variance with the $ab$-plane spiral, here we deal with an exact compensation of A- and B-contributions to the net electric polarisation $\left\langle P_a(y)\right\rangle$: $\left\langle P_a(y)\right\rangle _A=\left\langle P_a(y)\right\rangle _B =0$.


For the $ac$-plane spiral the molecular field ${\bf H}_0$ on the copper substituent is directed along the ${\bf b}$-axis, hence, in accordance with (\ref{Pa}) and in absence of external magnetic field $P_a(y)$ vanishes. The external magnetic field applied in $ac$-plane induces the oscillatory electric polarisation, however, the resultant polarisation $\left\langle P_a(y)\right\rangle$ vanishes irrespective of the field direction.

Thus the spin-spiral ordering in the CuO$_2$ chains of \LiV \,does induce due 
to the parity-breaking exchange a spontaneous electric polarisation on  
the Cu$^{2+}$ centers substituting for Li-ions resulting in a nonzero net 
ferroelectric polarisation  ${\bf P}\parallel$ $a$-axis only for $ab$-planar spin spirals, and a zero net effect both for $ac$- and $bc$-planar spin spirals.  
Such a behavior differs sharply from the 
predictions based on the spin current scenario (see Exp.(\ref{PM})), 
but it is strictly confirmed in experiments\cite{Naito,Naito1} 
where the ferroelectric polarisation in \LiV \, has been observed only along $a$-axis irrespective of the direction of the external magnetic field. Our model predicts the spontaneous electric polarisation  $P_a$ is proportional to the $m^2(T)$, $m(T)$   
being the magnetic order parameter, which agrees with the experimental findings\cite{Naito1}. The model reproduces  nicely  a 
smooth suppression of the effect with increasing
external field ${\bf h}\parallel {\bf c}$. Given reasonable estimates 
for the off-diagonal exchange integrals:  
$I_{b_{1g}e_{u}(\sigma)}\approx 0.1$ eV, and for the 
$g-u$ energy separation: $\Delta _{ug}\approx 4$ eV,
 we arrive at an estimate for the maximal value of the electric polarisation 
per unit cell occupied by an impurity
$P\approx 0.1e$\AA , or 
$P\approx 0.5\,\mu C/cm^{2}$\cite{remark2}. However, the actual net
polarisation is strongly reduced compared with this huge magnitude
due to the small impurity concentration $c \ll 1$
and the oscillatory character of $P_a(y)$.
Anyhow, even for reasonable small substitutions c(Cu$_{Li}$)$\approx 0.03-0.05$ 
\cite{Prokofiev,Kegler} and after averaging over impurity positions 
the 
net impurity polarisation in LiVCuO$_4$ can still be comparable
 with that observed in the gigantic multiferroic Ni$_3$V$_2$O$_8$\cite{Lawes}.


Above we assumed a weak coupling between the Cu$^{2+}$ substituent and the 
adjacent spin spirals treating the substituent 
as a paramagnetic impurity. However, the Cu$^{2+}$ substituent itself can 
locally distort   the spin spirals. Irrespective of the sign of the exchange parameters the Cu$^{2+}$ substituent - spin spiral exchange interaction 
favours locally a ferromagnetic inter-chain coupling with a phase shift $\Delta \alpha =0$, thus leading to a 
inter-chain coupling competing with that of the ideal chain system. 
 The role played by this effect in magnetoelectric coupling deserves  
a special investigation.
Also the effect of an external magnetic field on the ferroelectricity 
in spiral magnets seems to be more involved than it appears 
in our simplified model theory, and 
needs both in a more detailed and sophisticated description of interacting impurity centers, spirals, and a field induced
changes of the
spirals themselves.
In particular,  the pecularities near 8T observed
in the magnetization measurements \cite{Drechsler,Blanks} may evidence a breaking of the weak Cu$^{2+}$ substituent - spin spiral coupling approximation thus making a
breakdown of the ferroelectricity observed near 8T \cite{Naito,Naito1} quite natural.  
Finally, also the effect of finite temperatures should be considered. 
As regards the quantitative estimates of the net electric polarisation in \LiV \, it is worth noting that, strictly speaking, the parity-breaking exchange results in electric polarisation of both exchange-coupled centers with a trend toward  a cancellation, if the inversion center takes place inbetween. Indeed, the Cu$^{2+}$ substituent will polarise the four adjacent CuO$_4$ centers which net polarisation, however, fails to cancel $P_a(y)$ due to a  nonequivalence of the exchange-coupled CuO$_4$ centers. 

Anyhow, we have shown that in contrast to a spin-current scenario our  
simplified microscopic model already catches qualitatively correct the main features of the multiferroicity observed in \LiV, namely the  $a$-axis direction of the net electric polarisation for the $ab$-plane spin spiral ordering and its vanishing for the $bc$- and $ac$-planar  spirals. Along with  a reasonable estimation of the polarization,  the model is believed to explain the subtleties  of the field dependence, in particular, the suppression of the polarisation with increasing external magnetic field $\parallel$ $c$-axis.

In conclusion, we have discussed  
recent observations of 
the multiferroic behavior in the 1D chain cuprate LiVCuO$_4$  
with the edge-shared 
CuO$_4$ plaquettes and incommensurate spin ordering 
 and argued that its peculiarities can be consistently explained if one takes into 
account the exchange-induced electric polarisation on the Cu$^{2+}$ centers 
substituting for Li-ions in LiVCuO$_4$. Thus we deal with an unusual system where in sharp contrast with manganites the multiferroicity happens in two different subsystems, that is magnetic spirals induce electric polarisation at Cu$^{2+}$ ions substituting for native Li$^+$ ions. These substituent centers are 
proved to be an effective probe of the spin incommensurability and 
magnetic field effects.
The experimental investigation of possible relations between physical properties 
and the extent of nonstoichiometry as well as the theoretical 
consideration of analogous scenarios in other multiferroics would be of 
considerable scientific and potential practical interest.

{\it Note added.} After the submission of the manuscript, we became aware of an independent experimental study of the multiferroicity in \LiV \, by Schrettle {\it et al.}\cite{Schrettle}. In contrast with  Yasui  {\it et al.}\cite{Naito1}, the authors have found a weak $c$-axis electric polarization also for a $bc$-plane spin spiral. Generally speaking, this discrepancy does not necessarily contradict the nonstoichiometric nature of the multiferroic behaviour which makes it strongly sample dependent. In the general framework of an impurity model this might point to a manifestation of  weak charge transfer effects between the Cu$^{2+}$ substituent and the spin spirals. Anyhow,  further theoretical and experimental studies are needed to uncover all the subtle details of the unusual and challenging  multiferroicity in \LiV.

\acknowledgments
We thank Yu. Panov, A. Loidl, U.\ R\"o{\ss}ler, K. D\"orr, R. Kuzian and H. Rosner for discussions.
The  DFG and RFBR Grants (Nos.  06-02-17242, 06-03-90893,  and  07-02-96047)
 are acknowledged for financial support.


\begin{thebibliography}{0}


\bibitem{Naito}
\Name{ Naito Y.\ {\it et al.}}
  \REVIEW{J.\ Phys.\ Soc.\ Jpn.}{76}{2007}{023708}

\bibitem{Naito1}
\Name{Yasui Y.\ {\it et al.}}
  \REVIEW{arXiv:cond-mat/0711.1204}{}{2007}{}

 
\bibitem{Cheong}
\Name{ Park S.\ {\it et al.}}
  \REVIEW{Phys. Rev. Lett.}{98}{2007}{057601}
  
 \bibitem{Katsura1}
\Name{Katsura H.\ {\it et al.}}
  \REVIEW{Phys. Rev. Lett.}{95}{2005}{057205}
   
\bibitem{Mostovoy}
\Name{Mostovoy M.}
  \REVIEW{Phys.\ Rev.\ Lett.}{96}{2006}{067601}
 
\bibitem{Lawes}
\Name{ Lawes G.\ {\it et al.}}
  \REVIEW{Phys. Rev. Lett.}{95}{2005}{087205}

\bibitem{Chapon}
\Name{ Chapon L.C.\ {\it et al.}}
  \REVIEW{Phys.\ Rev.\ Lett.}{96}{2006}{097601}

\bibitem{Betouras}
\Name{ Betouras Joseph J.\ {\it et al.}}
  \REVIEW{Phys.\ Rev.\ Lett.}{98}{2007}{257602}

\bibitem{Sergienko1}
\Name{Sergienko I.E.\ {\it et al.}}
  \REVIEW{Phys.\ Rev.\ Lett.}{97}{2006}{227204}
 
\bibitem{Sergienko}
\Name{Sergienko I.E. \and Dagotto E.}
  \REVIEW{Phys. Rev. B}{73}{2006}{094434}
  
\bibitem{TMS}
\Name{ Tanabe Y.\ {\it et al.}}
  \REVIEW{Phys.\ Rev.\ Lett.}{15}{1965}{1023}



\bibitem{Katsura2}
\Name{Tanaka S.\ {\it et al.}}
 \REVIEW{Phys. Rev. Lett.}{97}{2006}{116404}
   
\bibitem{Hu}
\Name{ Hu C.D.}
  \REVIEW{arXiv:cond-mat/0608470}{}{2006}{}     
       
\bibitem{Jia}
\Name{ Jia C.\ {\it et al.}}
  \REVIEW{arXiv:cond-mat/0701614}{}{2007}{}
  

\bibitem{CuFeAlO}
\Name{ Nakajima T.\ {\it et al.}}
  \REVIEW{J.Phys. Soc. Jap.}{76}{2007}{043709}

\bibitem{elsewhere}
\Name{Moskvin A.S. \and Drechsler S.-L.}
  \REVIEW{arXiv:cond-mat/0710.0496}{}{2007}{}  
  
\bibitem{DM-JETP}
\Name{Moskvin A.S.}
  \REVIEW{JETP}{104}{2007}{911}
  
\bibitem{DM-PRB}
\Name{Moskvin A.S.}
  \REVIEW{Phys.\ Rev.\ B}{75}{2007}{054505}     
  

\bibitem{V-Cu}
We note that \LiV \, is not a vanadate but a lithium vanadyl cuprate. For this reason we prefer the short notation \LiV \, instead of the standard one LiCuVO$_4$.

\bibitem{Prokofiev}
\Name{Prokofiev A.V.\ {\it et al.}}
  \REVIEW{Journal of Solid State Chemistry}{177}{2004}{3131}

\bibitem{Brown}Brown I.D. and D.\ Altermatt, 
\REVIEW{Acta Crystallogr.\ Sect.\ B: Struct.\ Sc.}{41}{1985}{244} 

\bibitem{Gibson}
\Name{Gibson B.J.\ {\it et al.}}
  \REVIEW{Physica B}{350}{2004}{e253}

\bibitem{Enderle}
\Name{ Enderle M.\ {\it et al.}}
  \REVIEW{Europhys.\ Lett.}{70}{2005}{237}

\bibitem{Butgen}
\Name{B\"{u}ttgen N.\ {\it et al.}}
  \REVIEW{ArXiv:cond-mat/0701083v1}{}{2007}{}

\bibitem{Kegler}
\Name{Kegler C.\ {\it et al.}}
  \REVIEW{Phys. Rev. B}{73}{2006}{104418}

\bibitem{remark2}
A partial cancellation ($\sim 60\%$) should be made as actually the
  $b_{2g}$ orbital is a superposition of Cu 3$d$ and O 2$p$ contributions.  
  
 \bibitem{Drechsler}
\Name{ Drechsler S.-L.\ {\it et al.}}
\REVIEW{J. of Magnet. \& Magnet.\ Mat.}{316}{2007}{306}
\bibitem{Blanks}
\Name{ Blanks M.\ {\it et al.}}
\REVIEW{J.\ Phys.-Cond.\ Mat.}{19}{2007 }{145227} 
  
\bibitem{Schrettle}
\Name{ Schrettle F.\ {\it et al.}}
  \REVIEW{arXiv:cond-mat/0712.3583v1}{}{2007}{}



\end{thebibliography}
\end{document}